\documentclass[twocolumn,aps,pra,showpacs,pdftex,
fixfloats]{revtex4-1}
\usepackage{bm}
\usepackage{dcolumn}
\usepackage{graphicx}
\begin{document}
\newcommand{\PRA}{\textit{Phys. Rev.} A}
\newcommand{\bit}{\bibitem}
\newcommand{\tbf}{\textbf}
\newcommand{\tit}{\textit}
\newcommand{\JPCS}{\textit{J. Phys.} C: Conf. Ser.}
\newcommand{\JPB}{\textit{J. Phys.} B}
\newcommand{\etal}{\textit et. al.}

\title{Photoionization of atoms and ions from endohedral anions}
\author{ V. K. Dolmatov}
\email[Send e-mail to:\ ]{vkdolmatov@una.edu}
\address{Department of Chemistry and Physics, University of North Alabama,
Florence, Alabama 35632, USA}

\author{L. V. Chernysheva}
\address{Ioffe Institute, 194021 St.Petresburg, Russia, larissa.chernysheva@mail.ioffe.ru}

\author{V. G. Yarzhemsky}
\address{Kurnakov Institute of General and Inorganic Chemistry of RAS, 119991, Moscow, Russia; vgyar@igic.ras.ru}

\begin{abstract}
We study the interconnection between the results of two qualitatively different approximate calculations of photoionization cross sections, $\sigma_{n\ell}$, for neutral atoms ($A$) or their cations ($A^+$), centrally confined inside a fullerene-anion shell, $C_{N}^{q}$ , where $q$ represents the negative excess charge on the shell. One of the approximations, frequently employed in previous studies, assumes a uniform excess negative charge distribution over the entire fullerene shell, by analogy with a charged metallic sphere. The other approximation, not previously discussed in the literature, considers the quantum states of the excess electrons on the shell, determined by specific $n$ and $\ell$ values of their quantum numbers. Remarkably, both methods yield photoionization cross sections for the encapsulated species that are close to each other. Consequently, we find that the photoionization of the encapsulated atoms or cations inside a $C_{N}^{q}$ anion is minimally influenced by the quantum states of the excess electrons on the fullerene shell. Furthermore, we demonstrate that the aforementioned influence decreases even further with an increasing size of the confining fullerene shell. All this holds true at least under the assumption that the confined atom or cation is compact, i.e., its electron density remains primarily within itself rather than being drawn into the fullerene shell.
This remarkable finding results from Hartree-Fock calculations combined with  a popular modeling of the fullerene shell, where it is modeled by an attractive spherical annular potential.
\end{abstract}

\maketitle
\section{Introduction}

Among various theoretical studies on photoionization of atoms, $A$, and ions (cations), $A^{+}$, confined inside a negatively or positively charged  fullerene shell, C$_{N}^{q}$ ($q$ is a charge on the fullerene cage: $q= \pm |q|$ and $N$ is the number of carbon atoms making the fullerene cage/shell; throughout the rest of the paper, we will use the terms ``shell'' and ``cage'' interchangeably), a noticeable body of research results
has been obtained within the framework of a simple semi-empirical model
\cite{DolmMansPRA2006,DolmCravenPRA2009,LudlowJPB2010,KumarVarmaJPCS2015,KumarVarmaPRA2016,EPJD2016,JChemPhys2017,VarmaPRA2020,VarmaJPB2021,VarmaPhysSCR2021,DubeyJPB2021,CJP2022,VarmaPRA2023}.
The key assumption inherent in the model is that the excess charge $q$ on C$_{N}^{q}$ is uniformly spread over the entire outer surface of the fullerene cage in a direct analogy to a metallic sphere.

In the present paper, we investigate the validity of such assumption in relation to the photoionization process of a neutral atom, $A$, or its cation, $A^+$, confined inside a fullerene \textit{anion}, designated as
$A@{\rm C}_{N}^{q}$ or $A^{+}@{\rm C}_{N}^{q}$ ($q < 0$), respectively.

There are good reasons for such study. Firstly, said approximation neglects to account for the actual quantum state(s) of the excess electron(s) on the fullerene cage, such
as the electrons' principal, $n$, and orbital, $\ell$, quantum numbers. Secondly, in that approximation, the electron density of the excess electron(s) on the shell is uniformly distributed, as has already been mentioned, exclusively along the outer surface of the fullerene shell. Meanwhile, it is known, see, e.g., \cite{Lohr,ABK98,Dolm2020,Dolm2022} that the excess electron density does spread inside the hollow interior of the fullerene as well. Consequently, the question arises: how will the calculation results obtained within the former model change if one takes into account both the quantum states and the spread of the electron density of excess electrons from the shell into its interior? To the best of our knowledge, such study has not been conducted to date, leaving the question open. We address this question by providing the answer obtained from the research described in the present paper.

To meet the goal, and also to ensure the adequacy for comparing results obtained in both the framework of the former model and the improved model, suggested in the present paper, we retain the same semi-empirical modeling of a neutral C$_{N}$ shell as in the aforementioned previous studies. However, instead of uniformly spreading the excess charge over the neutral shell, we suggest a way to account for the quantum states of the excess charge in $A$@C$_{N}^{q}$ [here and throughout the paper, $A$ stands for either a neutral atom or a positive  ion (cation), for brevity].

Surprisingly, we find that the photoionization of the encapsulated atoms or cations inside a C$_{N}^q$ anion is minimally influenced by the quantum states of the excess electrons on the fullerene shell. Moreover, we demonstrate that said influence decreases with an increasing size of the confining fullerene shell, from C$_{60}$ to the giant C$_{240}$, in our study. Our findings hold true at least under the assumption that the confined atom is compact, i.e., its electron density remains primarily within the atom itself rather than being noticeably drawn into the fullerene shell.

We choose the hydrogen atom, H,  as well as the He atom and its ion He$^{+}$ as species to be confined inside a fullerene anion, as a case study.

Atomic system of units (a.u.) ($|e|=m_{\rm e}=\hbar = 1$, where $e$ and $m_{\rm e}$ are the electron's charge and mass, respectively, is used throughout the paper unless specified otherwise.

\section{Theory}
\subsection{Review of the former modeling of $A$@C$_{N}^{q}$}

The quintessence of the model is as follows.

Firstly, a neutral C$_{N}$ cage is modeled by a $U_{\rm C}(r)$ spherical annular, i.e., rectangular in the radial coordinate potential of  a certain  inner radius, $r_{0}$, thickness, $\Delta$, and depth, $U_{0}$:
\begin{eqnarray}
U_{\rm C}(r) = \left\{
\begin{array}{ll}
-U_{0}, & \mbox{if $r_{0} \le r \le r_{0} + \Delta$} \\
0, & \mbox{otherwise.}
\end{array}
\right.
\label{SWP}
\end{eqnarray}

Such modeling of a C$_{N}$ cage was apparently first proposed by Puska and Nieminen \cite{PuskaPRA93}.  Later, it was widely used in numerous theoretical studies on photoionization and other elementary processes involving endohedral fullerenes (also referred to as endohedral atoms/ions in this paper), $A$@C$_{N}$. The reader is referred to the works cited above, as well as, for example, to a relatively recent review \cite{DeshmukhEPJD2021} on these topics).

Secondly, following the work \cite{DolmMansPRA2006}, the effect of the charged shell C$_{N}^{q}$ on elementary atomic processes involving fullerene anions, $A$@C$_{N}^{q < 0} $, is taken into account by a uniform distribution of the excess charge $q$ over the entire
outer surface of the C$_{N}$ cage.  This leads to the appearance of an additional Coulomb spherical potential, $V_{q}(r)$, both inside and outside the fullerene cage:
\begin{eqnarray}
V_{q}(r) = \left\{
\begin{array}{ll}
\frac{q}{r_0 + \Delta}, & \mbox{if $ 0 \le r \le r_{0} + \Delta$} \\
\frac{q}{r}, & \mbox{otherwise.}
\end{array}
\right.
\label{eqVz}
\end{eqnarray} 

Consequently, the entire model potential, $U_{\rm C}^q(r)$,
 for a charged C$_{N}^{q}$ cage, is taken as the sum of the two potentials: $U_{\rm C}(r)$ [Equation~(\ref{SWP})], and $V_{q}(r)$ [Equation~(\ref{eqVz})]. That is,
\begin{eqnarray}
U_{\rm C}^q(r)= U_{\rm C}(r)+ V_{q}(r).
\label{eqV}
\end{eqnarray}

Next, to address the calculation of the structure and spectra of the encaged atom/ion, $A$, in the endohedral fullerene anion, $A$@C$_{N}^{q}$, the $V_q(r)$ potential is added to the Shr\"{o}dinger equation (for a single-electron atom), or to the Hartree-Fock, or the relativistic Dirac-Fock equations, or to the equations of other kinds of approximations for a multielectron atom.

For instance, the thus modified non-relativistic Hartree-Fock equation, applicable to $A$@C$_{N}^{q}$, is as follows:
\begin{eqnarray}
&&\left[ -\frac{\Delta}{2} - \frac{Z}{r} +U_{\rm C}^q(r) \right]\psi_{i}
({\vec r}) + \sum_{j=1}^{N_A} \int{\frac{\psi^{*}_{j}({\vec r}')}{|{\vec
r}-{\vec r}'|}} \nonumber \\
 && \times[\psi_{j}({\vec r}')\psi_{i}({\vec r})
- \psi_{i}({\vec r}')\psi_{j}({\vec r})]d {\vec r}' =
\epsilon_{i}\psi_{i}({\vec r}). \label{eqHF}
\end{eqnarray}
Here, $N_A$ is the number of electrons in the multielectron atom $A$, $i$ (and, similar,  $j$) denotes the set  of the principal ($n_i$), orbital ($\ell_i$), magnetic ($m_{\ell_i}$) and spin-magnetic ($m_{s_i}$) quantum numbers for the $i$-th atomic electron in  $A$@C$_{N}^{q}$. For the continuum spectrum of the $i$'s electron, $\epsilon_i$ is to be replaced by the energy of the continuum spectrum,  $\epsilon$.
Furthermore, the function $\psi_{n_i \ell_i m_{\ell_i}}({\vec r})$ of the $i$'s electron admits a separable representation: $\psi_{n_i \ell_i m_{\ell_i}}({\vec r})=r^{-1} P_{n_i\ell_i}(r)Y_{\ell_i m_{\ell_i}}(\theta, \phi)$.

We will frequently refer to such model of the fullerene anion as ``structureless'' model or approximation in this paper, to underline that it ignores possible quantum states of the excess electronic charge on the cage.

Next, the dipole photoionization cross section, $\sigma_{n\ell}$,  of a $n\ell$-subshell of the encapsulated atom is calculated using the standard formula (see, e.g., \cite{Amusia_book}):
\begin{eqnarray}
\sigma_{n\ell}(\omega)= \frac{4\pi{^2}\alpha N_{n\ell}}{3(2l+1)} \omega [l
D_{\ell-1}^{2}+ (\ell+1)D_{\ell+1}^{2}].
\label{eqCC}
\end{eqnarray}
Here, $\omega$ is the photon energy, $\alpha$ is the fine structure constant, $N_{n\ell}$ is the
number of electrons in the $n\ell$ subshell, $D_{l\pm 1}$ is a radial dipole
photoionization amplitude,
\begin{eqnarray}
D_{\ell\pm 1}=\int_{0}^{\infty}{P_{\epsilon \ell\pm 1}(r) r P_{n\ell}(r) dr}.
\label{eqD}
\end{eqnarray}
\subsection{Accounting for quantum states of the excess electrons in $A$@C$_{N}^{q}$}

In our modeling of A@C$_{N}^q$, the extra electron(s) in the A@C$_{N}^q$ fullerene anion is(are) bound by the central field which is the sum of the fields of the atom and neutral C$_{N}$ cage. Therefore, the excess electronic states are characterized by quantum numbers $n$ and $\ell$. Let us designate such endohedral fullerene anions as A@C$_{N}^q(n\ell)$ or A@C$_{N}^q(n\ell, n'\ell')$ and so on, depending on how many excess electrons exist and how many different $n\ell$-states they occupy in the fullerene anion. For example, He@C$_{N}^{-1}(2s)$ means that the single attached electron resides in a $2s$ state on the singly-charged fullerene shell, He@C$_{N}^{-2}(2s2p)$ indicates  that one of the two attached electrons in the doubly-charged  anion is in a $2s$ state and the other electron is in a $2p$ state, and so on.

To account for the $n\ell$-structure of the excess electronic charge in the anion and, at the same time, retain the spirit of the semi-empirical framework for modeling the fullerene cage, we follow the methodology outlined in \cite{Dolm2020}. The essence of the latter is as follows.

 The A@C$_{N}^q(n\ell,...)$ system is a complete (or ``single'') system in the sense that the state(s) of the attached  electron(s) is(are) affected by the field of the encapsulated atom, whereas the states of the atomic electrons, in turn, are affected by both the field of the cage itself, $U_{\rm C}(r)$, and by the field(s) of the attached electron(s). To account for the mutuality of the states of the atomic electrons and those of the excess electrons on the fullerene cage, we simply solve, simultaneously, a system of the $N_{tot}$ Hartree-Fock equations for the \textit{``atom + fullerene anion''} system, where $N_{tot}=N_A+N_q$, with $N_q$ being the number of the excess electrons on the fullerene cage, and $N_A$ the number of the electrons in the encapsulated atom.
\begin{eqnarray}
&&\left[ -\frac{\Delta}{2} - \frac{Z}{r} +U_{\rm C}(r) \right]\psi_{i}
({\vec r}) + \sum_{j=1}^{N_{\rm tot}} \int{\frac{\psi^{*}_{j}({\vec r}')}{|{\bm
r}-{\vec r}'|}} \nonumber \\
 && \times[\psi_{j}({\vec r}')\psi_{i}({\vec r})
- \psi_{i}({\vec r}')\psi_{j}({\vec r})]d {\vec r}' =
\epsilon_{i}\psi_{i}({\vec r}). \label{eqHF'}
\end{eqnarray}
Here, $i$ and $j$ now run from unity to $N_{\rm tot}$: $i, j = 1, ..., N_{\rm tot}$.
This system of equations differs from  Equation~(\ref{eqHF}) in that the $N$ number is replaced by $N_{\rm tot}$ and the $U_{\rm C}^{q}$ potential is replaced by $U_{\rm C}$, [Equation~(\ref{SWP})]. It allows one to calculate the needed energies, $\epsilon_i$, and the wavefunctions, $\psi_{i}({\vec r})$, of both the atomic and excess electrons in A@C$_{N}^q(n\ell,...)$ and apply them to calculations of the photoionization cross sections of the thus confined atom A.

We will frequently refer to such model of the fullerene anion as ``structured'' model or approximation in this paper, to underline that it accounts for possible quantum states of the excess electronic charge on the cage.

Now, however, a complication arises due to the existence of various total terms for the A@C$_{N}^q(n\ell,...)$ system, which is now an open-shell system. In the present study, we bypass this complication by utilizing the term-average Hartree-Fock formalism \cite{YarzhCher2024} to calculate the energies and wavefunctions of both the atomic and excess electrons in the A@C$_{N}^q(n\ell,...)$ system. To avoid diverting the reader's attention from the main topic of the paper, a review of the term-average Hartree-Fock formalism is provided in Appendix.

Lastly, in performed calculations for the present study, we used the values for $r_0$, $\Delta$ and $U_0$ for C$_{60}$/C$_{240}$ as stated and discussed in \cite{DolmBrMans2008}:
$r_0\approx 5.8/12.6$, $\Delta  \approx 1.9/1.9$, $U_0 \approx 0.302/0.378$ a.u., respectively.

\section{Results and Discussion}
\subsection{${\rm H}(1s)@{\rm C}_{N}^{-1}$ versus ${\rm H}(1s)@{\rm C}_{N}^{-1}(n\ell)$}
Here, we present and discuss calculated results concerning the photoionization of the ground-state hydrogen atom, H$(1s)$, encapsulated inside the singly-charged fullerene anion ($q=-1$). These results were obtained in the framework of both  aforementioned models for the endohedral fullerene anions - the structureless [${\rm H}(1s)@{\rm C}_{N}^{q}$] and structured [${\rm H}(1s)@{\rm C}_{N}^q(n\ell)$] models.
 As case studies, we arbitrarily choose $n\ell = 2p$ and $3d$, and we run calculations for C$_{N}$ fullerene cages with both $N=60$ and $N=240$ carbon atoms in the cage, respectively.
%
%The notation ${\rm H}(1s)@{\rm C}_{N}^{-1}$ underlines, that the hydrogen atom is confined inside a fullerene anion where the excess charge $q=-1$ (the charge of one electron) is uniformly distributed over the entire outer surface %of the fullerene shell. Alternatively, the notation ${\rm H}(1s)@{\rm C}_{N}^{-1}(n\ell)$ stands for the case of the hydrogen atom encapsulated into a singly charged fullerene shell where the excess electron in the shell resides %in a certain $n\ell$-state which we take arbitrarily as a $2p$ state or a $3d$ state in our study.

In Figure $1$, the $P_{1s}$ radial function of the encapsulated H atom is depicted. Additionally, the $P_{2p}(r)$ and $P_{3d}(r)$ radial functions are depicted for the $2p$ and $3d$ orbitals of the attached electron, respectively, within the endohedral fullerene anions.

\begin{figure}
\includegraphics[width=8cm]{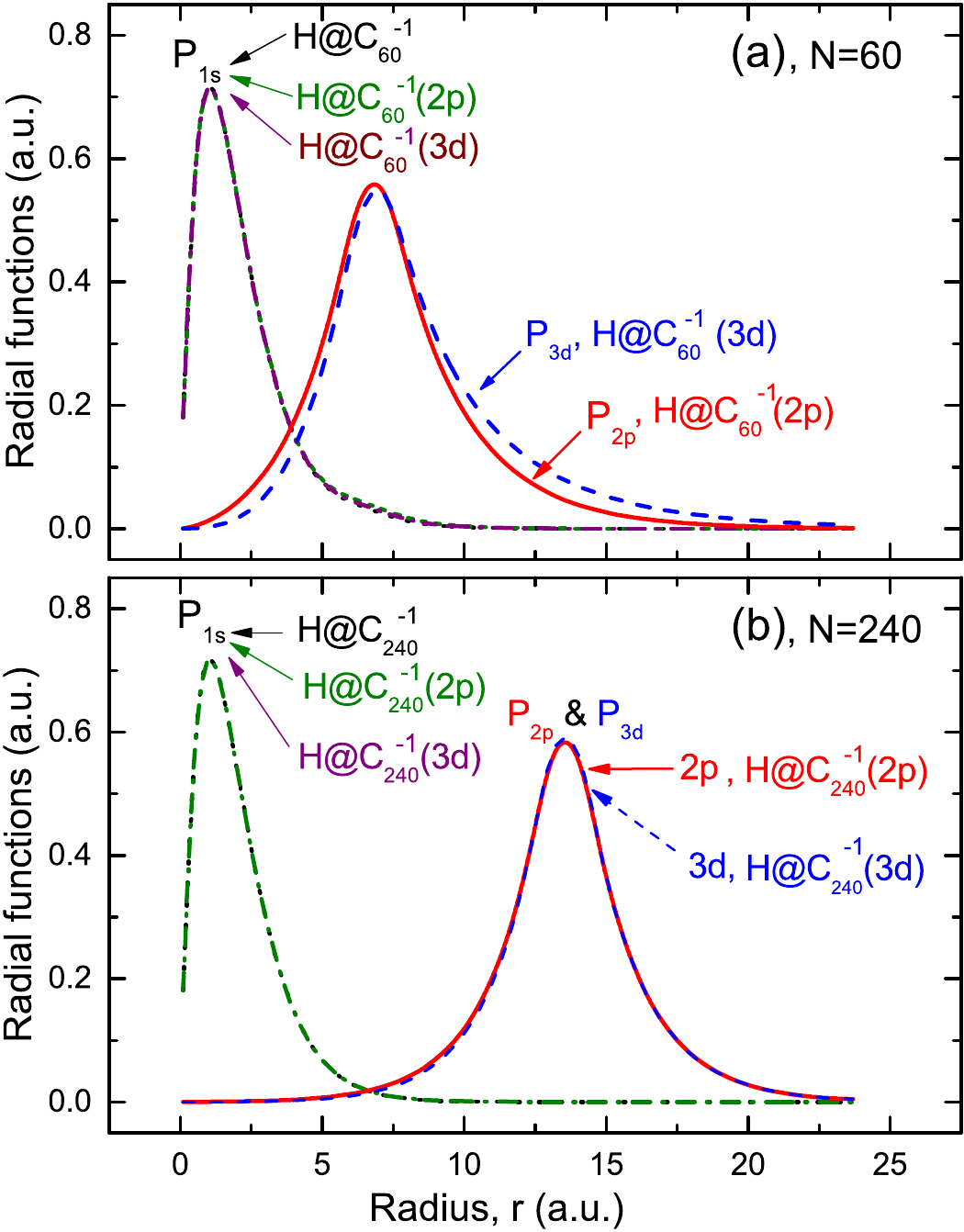}
\caption{(Color online) (a) Calculated $P_{1s}(r)$ radial functions for ${\rm H}@{\rm C}_{60}^{-1}$ as well as the $P_{1s}(r)$, $P_{2p}(r)$ and $P_{3d}(r)$ electronic radial functions for ${\rm H}@{\rm C}_{60}^{-1}(n\ell)$ with $n\ell= 2p$ and $n\ell = 3d$ (see text), as designated in the figure. (b) Calculated $P_{1s}(r)$ radial functions for ${\rm H}@{\rm C}_{240}^{-1}$ as well as the $P_{1s}(r)$, $P_{2p}(r)$ and $P_{3d}(r)$ electronic radial functions for ${\rm H}@{\rm C}_{240}^{-1}(n\ell)$ with $n\ell= 2p$ and $n\ell = 3d$ (see text), as designated in the figure.
Note, the $P_{1s}$ function for the encapsulated hydrogen atom depends so little on both the presence and state of the excess electron in the fullerene shell in all calculations that all graphs for $P_{1s}$ are practically
totally blended with each other in the figure.
}
\label{fig1}
\end{figure}

Firstly, it can be concluded from Figure~\ref{fig1} that the $1s$ electron density in the encapsulated atom remains largely unaffected by the excess charge on the fullerene cage, especially in the case of the C$_{240}$ cage.

Secondly, note how the $P_{2p}(r)$ and $P_{3d}(r)$ functions of the excess electron extend, to some degree, into the hollow interior of the C$_{N}$ shell, thereby overlapping with the $P_{1s}$ function of the encapsulated H atom. This, obviously, should make the $1s$-photoionization
cross section, $\sigma_{1s}$, of the  ${\rm H}(1s)@{\rm C}_{N}^{-1}(n\ell)$ system differ from that of the structureless  ${\rm H}(1s)@{\rm C}_{N}^{-1}$ system. How strong can the difference be? One of the aims of the present study is to answer this question.

Thirdly, one can see that the overlap of the $P_{1s}$ function with the functions for the $2p$ and $3d$ excess electrons is far less significant in the case associated with a C$_{240}$ shell than in the other case. Hence, it is reasonable to expect that any differences in values of
$\sigma_{1s}$ between the ${\rm H}(1s)@{\rm C}_{N}^{-1}$ and ${\rm H}(1s)@{\rm C}_{N}^{-1}(n\ell)$ systems will diminish with the increasing size of the fullerene cage.

Fourthly, interestingly, the $2p$ and $3d$ functions of the excess electron differ insignificantly, especially in the case of the giant C$_{240}$ cage.
A similar situation was observed and explained for the $P_{n\ell}$ functions of the $n\ell$-electron attached to the empty fullerene cage \cite{Dolm2022}.
The reason for the noted indistinguishability between the functions with different $\ell$ values (the  $P_{2p}$ and $P_{3d}$ functions in our case) arises from the electrons' primary localization on the cage itself, which has a large radius. Consequently, the centrifugal barrier $\frac{\ell(\ell+1)}{2r^2}$, which induces the dependence of $P_{n\ell}$ on $\ell$, becomes so small that the difference between the functions with close $\ell$ values effectively vanishes. The latter is particularly true for the case of the giant C$_{240}$ cage.

Calculated $\sigma_{1s}$ photoionization cross sections of H@C$_{N}^{-1}$ and H@C$_{N}^{-1}(n\ell)$ with $n\ell = 2p$ and $3d$ are depicted in Figure $2$.
\begin{figure}
\includegraphics[width=8cm]{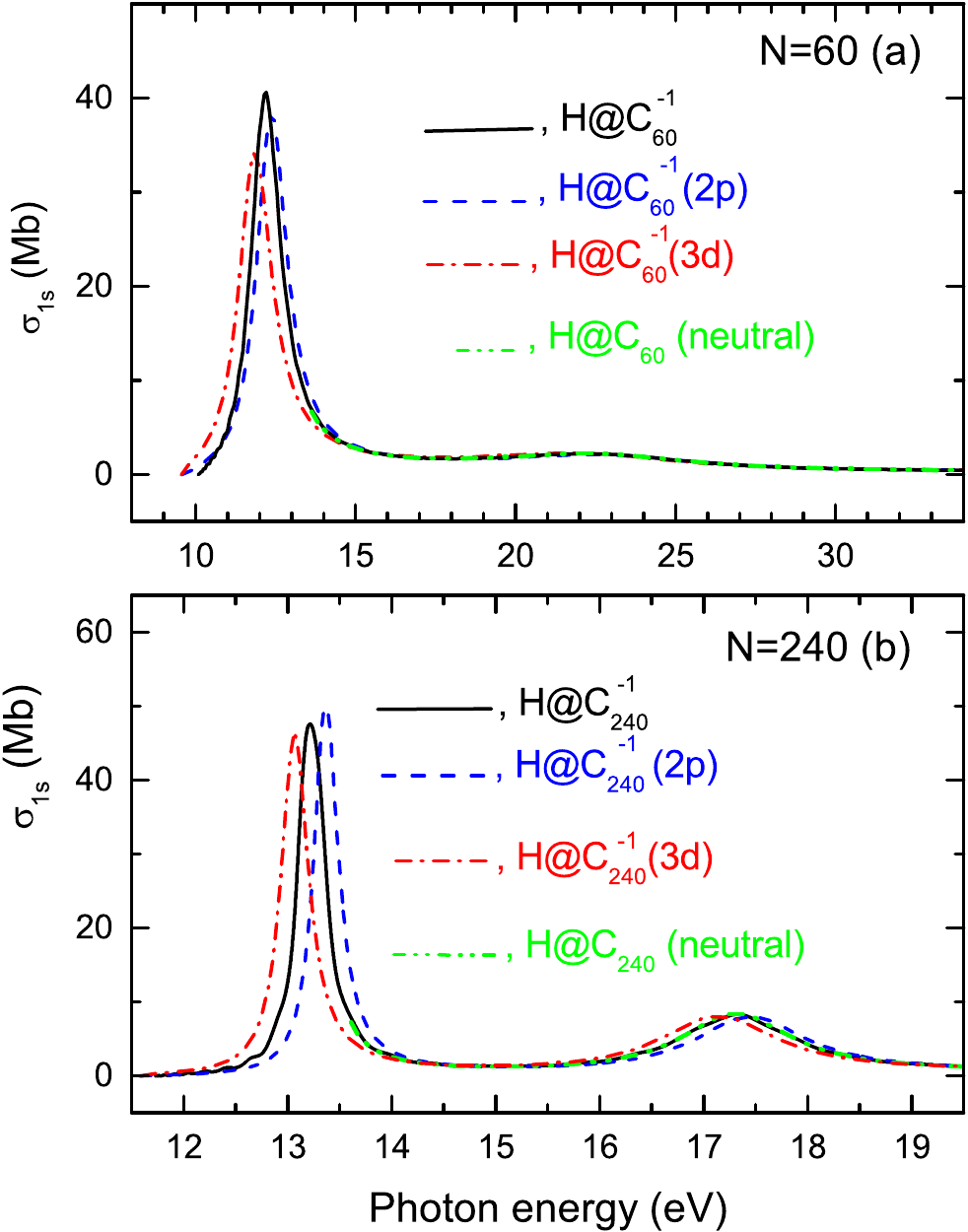}
\caption{(Color online) Calculated $\sigma_{1s}$ photoionization cross sections for the H$(1s)$ atom confined inside various fullerene anions: H@C$_{60}^{-1}$, H@C$_{60}^{-1}(2p)$, H@C$_{60}^{-1}(3d)$, H@C$_{240}^{-1}$ and H@C$_{240}^{-1}(2p)$, as designated in the figure.}
\label{fig2}
\end{figure}

One can see that some differences between these $\sigma_{1s}$ cross sections are seen only in the low-energy region. There, the sharp, intense maxima in $\sigma_{1s}$'s are somewhat shifted along the energy scale relative to each other without a significant change in their magnitudes. Thus, it is reasonable to summarize that the change in $\sigma_{1s}$'s between the H@C$_{60}^{-1}$, and H@C$_{N}(n\ell)$ systems are minimal. This reveals the first indication that the  photoionization cross section of the atom from an endohedral fullerene anion depends little on the state of the excess charge on the anion.

And, as expected, the differences between the discussed $\sigma_{1s}$'s are diminishing with the increasing size of the fullerene cage. This is obvious from comparing calculated data plotted in Figure $2$a (for $C_{60}$) with those plotted in Figure $2$b for the case of a C$_{240}$ cage. Note that a would-be larger shift in energy between the low-energy maxima in $\sigma_{1s}$'s, plotted in Figure $2$b, compared to the energy shift in Figure $2$a is deceptive. This is because the energy-scale range is less than $10$ eV in Figure $2$b, but is much broader, about $30$ eV, in Figure $2$a.

Finally, let us comment on the origin and nature of all the maxima in $\sigma_{1s}$'s displayed in Figure $2$. To meet the goal, we also plotted calculated $\sigma_{1s}$ photoionization cross sections of the hydrogen atom confined inside a neutral $C_{N}$ cage, designated as ``H$@{\rm C}_N$ (neutral)'' in the figure.

Firstly, the calculated $\sigma_{1s}$ values, plotted in Figure $2a$ or $2b$, are nearly identical down to about $14$ eV of photon energy. Secondly, for photon energies down to about $14$ eV, all $\sigma_{1s}$'s exhibit a maximum around $22$ eV for $N=60$ (though this maximum is poorly developed) and around $17$ eV for $N=240$. Therefore, the higher-energy maxima in $\sigma_{1s}$ for endohedral fullerene anions have the same origin as the maximum in $\sigma_{1s}$ for the neutral H@C$_N$ endohedral fullerene.
The origin of such maxima in the $\sigma_{n\ell}$ for atoms in neutral A@C$_N$ endohedral fullerenes has been extensively studied in previous works, e.g., \cite{PuskaPRA93,DeshmukhEPJD2021,Baltenkov,ConDolmMans,DolmAQC} (and references therein). These maxima arise from interference between the outgoing photoelectron wave from the encapsulated atom and the photoelectron waves reflected from the fullerene cage boundaries. In \cite{ConDolmMans}, these resonances were termed \textit{confinement resonances}. This term has become common in the literature, and we use it in the present paper as well. Thus, the observed higher-energy resonances in $\sigma_{1s}$ for H@C$_{60}^{-1}$, H@C$_{N}^{-1}(n\ell)$, and neutral H@C$_N$ are confinement resonances.

Next, in the photon energy region below about $14$ eV, all calculated $\sigma_{1s}$ values for fullerene anions show a sharp, strong resonance, whereas the $\sigma_{1s}$ for neutral H@C$_N$ does not extend into this region. Thus, these sharp lower-energy resonances originate from a different source compared to the confinement resonances.
Such resonances in the photoionization cross sections of atoms encapsulated within fullerene anions were predicted and studied in detail in \cite{DolmMansPRA2006}. It was demonstrated that these resonances arise due to an additional Coulomb potential barrier created by the excess charge on the fullerene cage. In \cite{DolmMansPRA2006}, these novel resonances in $\sigma_{n\ell}$ values were termed \textit{Coulomb confinement resonances}. To date, Coulomb confinement resonances have been explored in various contexts with different levels of detail in numerous theoretical studies (see, e.g., \cite{KumarVarmaJPCS2015,KumarVarmaPRA2016,VarmaPRA2020,VarmaPhysSCR2021,VarmaPRA2023} and references therein).
Therefore, the lower-energy resonance structures in $\sigma_{1s}$ for H@C${N}^{-1}$ and H@C${N}^{-1}(n\ell)$ are Coulomb confinement resonances.

To conclude, we have unraveled the first indication that the photoionization cross section of the atom within the endohedral fullerene anion is largely independent of the quantum structure of the excess charge on the anion. The most noticeable differences (albeit insignificant) in the photoionization cross sections between the structureless and structured fullerene anions primarily occur in the region of Coulomb confinement resonances.

\subsection{$({\rm He\ \&\ He^+)@C}_N^{-1}$, $({\rm He\ \&\ He^+)@C}_N^{-1}(n\ell)$ and $({\rm He\ \&\ He^+)@C}_N^{-2}(n\ell, n'\ell')$}

As another case study, we examine the photoionization of the He$(1s^2)$ atom and its ion, He$^+(1s^1)$, confined within the structureless fullerene anion cage, C$_{N}^q$, compared to when they are encapsulated inside the structured fullerene anion cages, ${\rm C}_{N}^q(n\ell)$ or ${\rm C}_{N}^q(n\ell, n'\ell')$. Specifically, we arbitrarily choose the He@C$_{N}^{-1}(2p)$ and He@C$_{N}^{-1}(3d)$ singly-charged  endohedral fullerene anions, as well as the He$^+@{\rm C}_N^{-2}(2s 2p)$ and He$^+@{\rm C}_N^{-2}(2s 3d)$ doubly-charged anions, with $N=60$ and $240$.

\begin{figure}
\includegraphics[width=8cm]{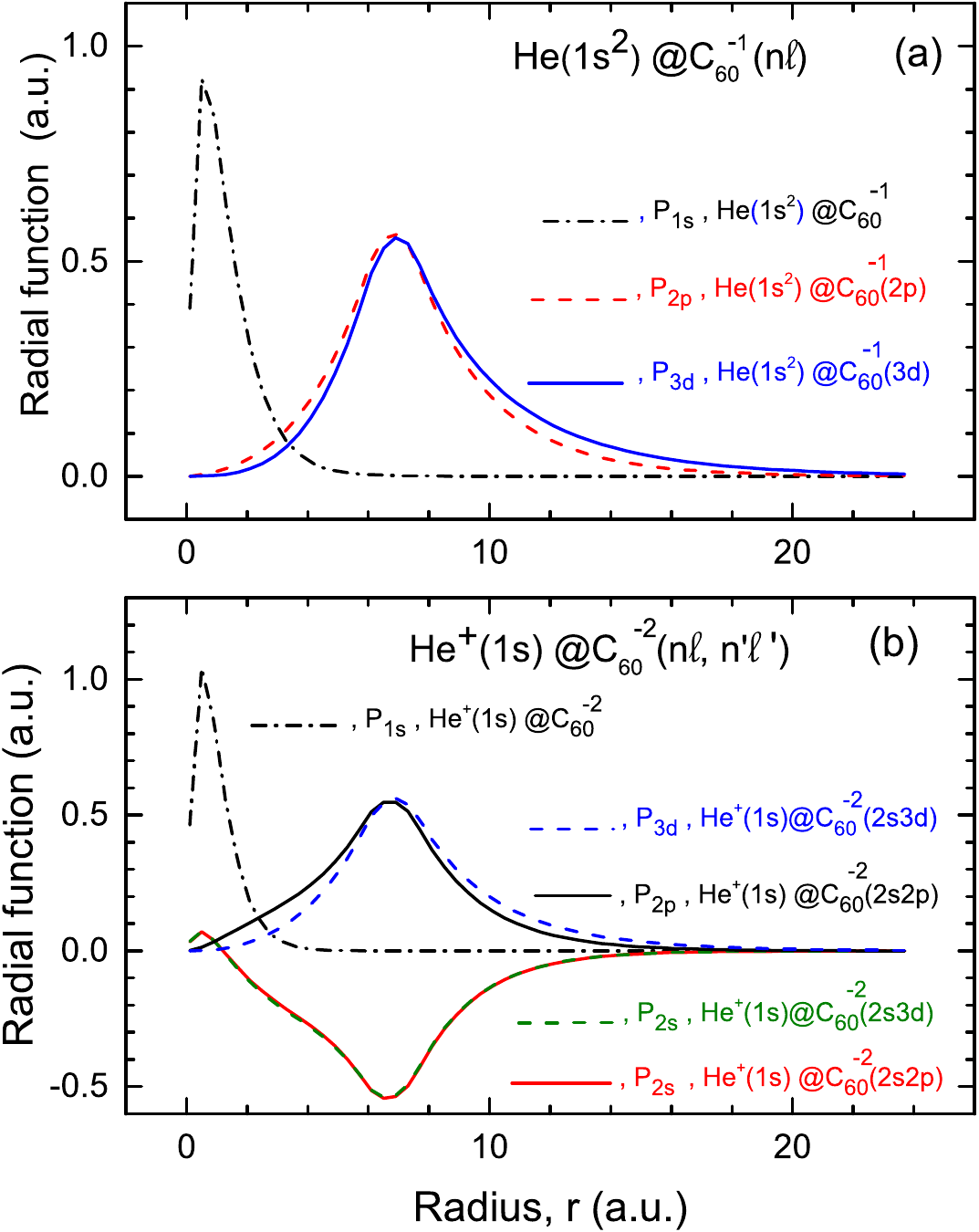}
\caption{(Color online) (a) Calculated $P_{1s}(r)$ of a neutral He$(1s^2)$ atom encapsulated inside structureless ${\rm C}_{60}^{-1}$, as well as the $P_{2p}$ and $P_{3d}$ functions of the attached electron in the structured He$@{\rm C}_{60}^{-1}(2p)$ and He$@{\rm C}_{60}^{-1}(3d)$ fullerene anions, as designated in the figure.
(b) Calculated $P_{1s}(r)$ of the He$^+$ ion encapsulated inside structureless ${\rm C}_{60}^{-2}$, as well as the $P_{2s}$, $P_{2p}$, and $P_{3d}$ functions of the attached electron in the corresponding structured He$^{+}@{\rm C}_{60}^{-2}(2s2p)$ and He$^{+}@{\rm C}_{60}^{-2}(2s3d)$ fullerene anions, as designated in the figure.
Note that, similar to the case of the H atom, the $P_{1s}(r)$ functions overlap with each other in each of the considered cases, so we plotted $P_{1s}(r)$ only for one of them in both parts of the figure.}
\label{fig3}
\end{figure}

The corresponding $P_{1s}$, $P_{2s}$, $P_{2p}$ and $P_{3d}$ electronic functions are plotted in Figure \ref{fig3}.

The behavior of the plotted functions follows the same trends as in the previously discussed case of the H atom inside the fullerene anion. Therefore, we believe that they are self-explanatory without new elements in the behavior, and we leave it to the reader to draw their own conclusions regarding these functions. It is important to stress, however, that, as earlier, there is overlap between the $P_{1s}$ atomic function and each of the corresponding $P_{2s}$, $P_{2p}$, and $P_{3d}$ functions. Accordingly, we explore below how this overlap between the electronic functions affects the $\sigma_{1s}$ photoionization cross section of He and He$^+$ from the structured fullerene anions compared to the structureless fullerene anions.

The correspondingly calculated  $\sigma_{1s}$ photoionization cross sections  are depicted in Figure $4$.

\begin{figure}
\includegraphics[width=8cm]{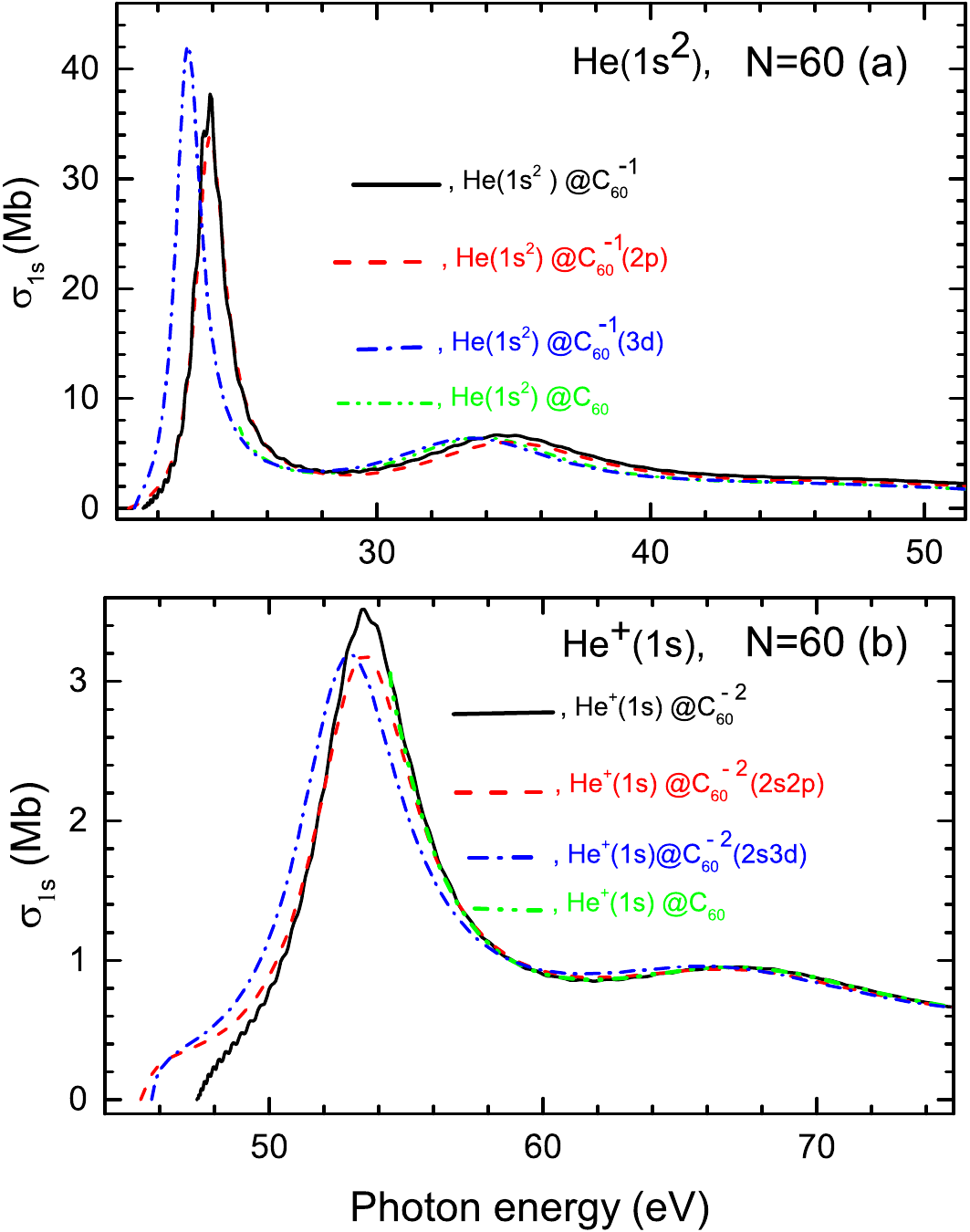}
\caption{(Color online) (a) Calculated $\sigma_{1s}$ photoionization cross sections for He$@{\rm C_{60}}^{-1}$ as well as for He$@{\rm C_{60}}(2p)$ and He$@{\rm C_{60}}(2p)$ along with $\sigma_{1s}$ for the helium atom confined inside a neutral $C_{60}$ shell, i.e., for
He$(1s^2)@C_{60}$, as designated in the figure.
(b) Calculated $\sigma_{1s}$ photoionization cross sections for He$^{+}@{\rm C_{60}}^{-2}$, He$^{+}@{\rm C_{60}}^{-2}(2s 2p)$ and He$^{+}@{\rm C_{60}}^{-2}(2s 3d)$, as well as $\sigma_{1s}$ He$^{+}$ inside a
neutral $C_{60}$ shell, i.e.,
He$^{+}@{\rm C_{60}}$, as designated in the figure.
}
\label{fig4}
\end{figure}

Firstly, similar to the hydrogen atom case discussed earlier, Figure \ref{fig4} reveals that $\sigma_{1s}$ for the encapsulated helium atom and its ion exhibit a pronounced, low-energy sharp resonance when photoionization occurs from the fullerene anions. This resonance is absent in the  photoionization  of He and He$^+$ encapsulated inside the neutral C$_{60}$ cage. Thus, as in the case of the hydrogen atom, these observed features are Coulomb confinement resonances. As photon energy increases, these Coulomb confinement resonances are succeeded by less pronounced resonances in both the charged and neutral endohedral fullerene systems. These latter features are, thus, ordinary confinement resonances.

Secondly, a key finding, however, is that the differences in $\sigma_{1s}$ between photoionization of He or He$^{+}$ from the structureless and various structured endohedral fullerene anions, respectively, are primarily observed in the Coulomb confinement resonances. However, these differences are relatively minor, with no significant qualitative or strong quantitative variations. We, thus, have uncovered one more indication that the photoionization cross sections of the atom or its ion confined inside the endohedral fullerene anion depend little on the quantum structure of the excess charge on the fullerene cage.

Next,  we present the results for calculated $\sigma_{1s}$ for He and He$^+$ confined within the structureless and various structured fullerene anions with a giant C$_{240}$ carbon cage. These calculated $\sigma_{1s}$'s are shown in Figure 5.

\begin{figure}
\includegraphics[width=8cm]{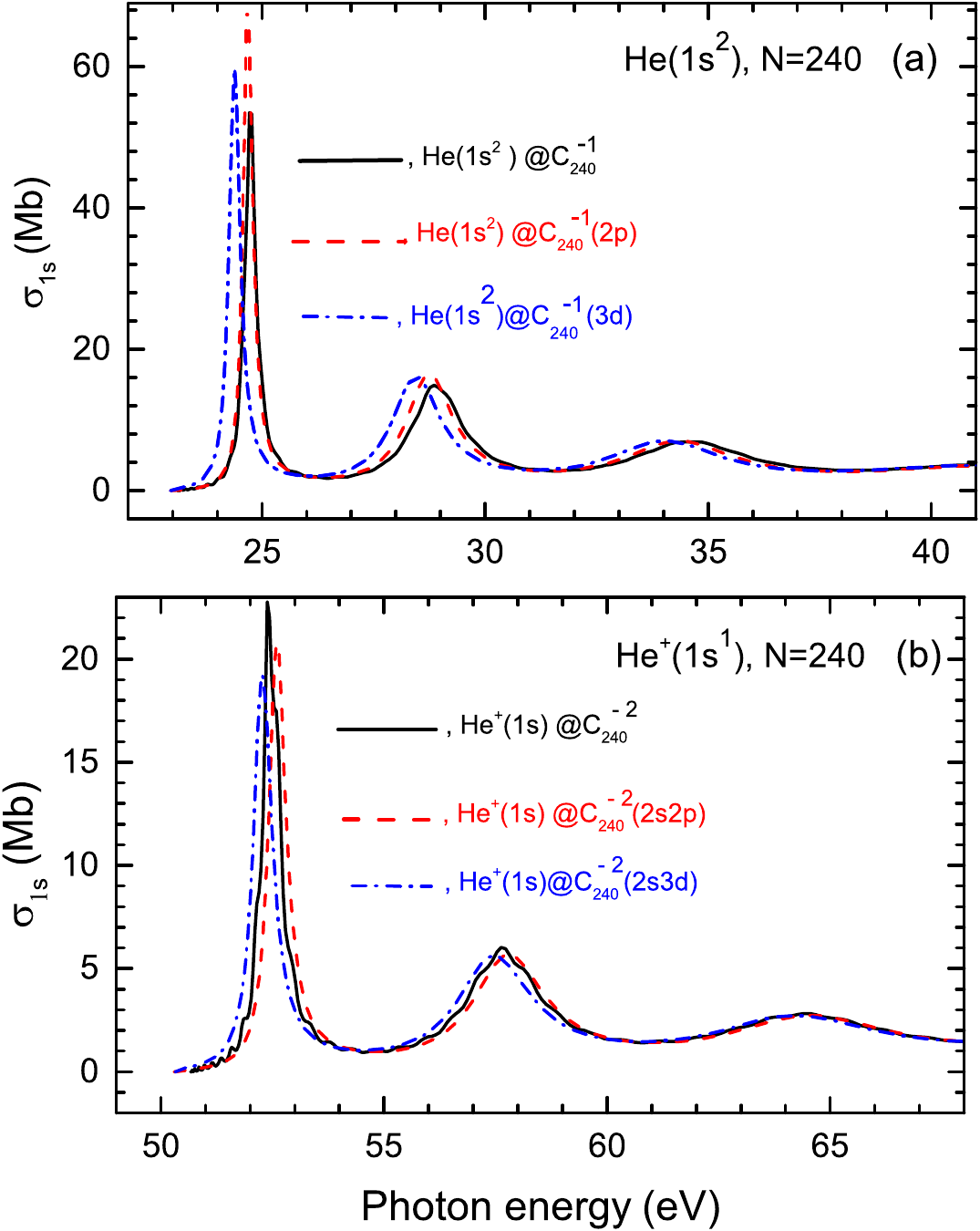}
\caption{(Color online) (a) Calculated $\sigma_{1s}$ photoionization cross sections for He$@{\rm C_{240}}^{-1}$ as well as for He$@{\rm C_{240}}(2p)$ and He$@{\rm C_{240}}(2p)$ along with $\sigma_{1s}$ for the helium atom confined inside a neutral $C_{240}$ shell, i.e., for
He$@{\rm C}_{240}$, as designated in the figure.
(b) Calculated $\sigma_{1s}$ photoionization cross sections for He$^{+}@{\rm C_{240}}^{-2}$, He$^{+}@{\rm C_{240}}^{-2}(2s 2p)$ and He$^{+}@{\rm C_{240}}^{-2}(2s 3d)$, as well as $\sigma_{1s}$ He$^{+}$ inside a
neutral $C_{240}$ shell, i.e.,
He$^{+}@{\rm C_{240}}$, as designated in the figure.
}
\label{fig5}
\end{figure}

The trends observed in the calculated $\sigma_{1s}$'s for the structureless and structured giant fullerene anions mirror those just discussed above for He and He$^+$ confined within the fullerene anions associated with the C$_{60}$ fullerene cage. It is worth noting, however, that the differences in calculated $\sigma_{1s}$'s are significantly smaller in giant fullerene cases compared to the previous ones, which was anticipated.

The results presented in this section  uncover that the photoionization cross sections of both the neutral atom and its ion encapsulated inside the fullerene anion cage are affected only little by the quantum structure of the excess charge on the cage.

\section{Conclusions}

In this study, we investigated the photoionization cross sections of compact H and He atoms, as well as the He$^{+}$ ion, encapsulated within a fullerene anion cage. The electron density of these encapsulated species remains unaffected by the fullerene cage.

The results and discussions indicate that the photoionization cross section of the centrally encapsulated atom or its cation inside the C$_{N}^q$ fullerene anion cage is minimally influenced by the quantum states of the excess electrons on the cage, provided the electron density of the encapsulated atom/ion remains predominantly within the atom/ion itself. This conclusion is likely extendable to other centrally confined compact atoms as well; we do not see why this would be untrue. The most noticeable differences in the calculated photoionization cross sections, which arise from considering the quantum states of the excess electrons on the fullerene cage, are primarily observed in the Coulomb confinement resonances. However, these differences are relatively small and diminish significantly with the increasing size of the fullerene cage. Consequently, both the original structureless model and the structured model of the fullerene anion yield photoionization cross sections that are in close agreement, especially for the case of the giant fullerene anion. Therefore, as one of the key findings of this research, we conclude that either model is equally effective for studying the photoionization cross sections of atoms encapsulated within the fullerene anion, as long as the electron density of the encapsulated atom/ion remains primarily within the atom/ion itself.

For atoms that significantly contribute their electron density to the fullerene cage, we anticipate that the photoionization cross sections of the inner electrons will depend insignificantly on the quantum structure of the excess electrons on the fullerene cage as well. However, this scenario involving such atoms requires a separate, independent study, which is beyond the scope of this paper.

\appendix
\section{Term-average Hartree-Fock approximation}

In this section, we outline the key steps that lead to the term-average methodology for the Hartree-Fock formalism  \cite{YarzhCher2024} (also with its extention to the calculation of many-body Feynman diagrams) used in the present study.

In the calculation of atomic wavefunctions using the Hartree-Fock method, the combined direct and exchange Coulomb interelectron interaction energy, $U_{\rm c}$, in the atom is given by:

%\begin{eqnarray}
%U_{c}=\sum\limits_{k}\left( \sum_{\lambda }F_{SL}{\lambda}R_{llll}_{\lambda }\right) +\sum\limits_{k\neq p}\left(\sum\limits_{\mu }F_{SLS'L'{L_{t}}{S_{t}}}^{\mu }R_{ll'll'}^{\mu}
% +\sum\limits_{k\neq p}\sum\limits_{\nu }G_{SLS'L'{L_{t}}{S_{t}}}^{\nu }R_{lll'l'}^{\nu }\right)
% \label{uc}
%\end{eqnarray}

\begin{eqnarray}
U_{c}=\sum\limits_{k}\sum_{\lambda }F_{SL}^{\lambda}R_{llll}^{\lambda}\nonumber \\
 +\sum\limits_{k\neq p}\sum\limits_{\mu }F_{SLS'L'{L_{t}}{S_{t}}}^{\mu }R_{ll'll'}^{\mu}\nonumber \\
 +\sum\limits_{k\neq p}\sum\limits_{\nu }G_{SLS'L'{L_{t}}{S_{t}}}^{\nu }R_{lll'l'}^{\nu }.
 \label{uc}
\end{eqnarray}

Here, the sum runs over all atomic subshells $k(p)$ and all  values of $\lambda
,\mu $ and $\nu $ that satisfy the triangular rule $\Delta \left( \ell\lambda
\ell\right) $, $\Delta \left( \ell\mu \ell\right) \cap \Delta \left( \ell^{\prime }\mu
\ell'\right) $ and $\Delta \left( \ell\nu \ell'\right) $,
provided the sum of the angular moments in brackets is even.

The Coulomb integral, expressed
in terms of the radial parts $P_{n_{k}\ell_{k}}$ of the atomic  wavefunctions, is given by the
standard formula:

\begin{eqnarray}
R_{\ell_{1}\ell_{2}\ell_{3}\ell_{4}}^{\lambda }=\int_{0}^{\infty }\int_{0}^{\infty }
\frac{r_{<}^{\lambda }}{r_{>}^{\lambda +1}}\nonumber \\
\times
P_{n_{1}\ell_{1}}(r)P_{n_{3}\ell_{3}}(r)
P_{n_{2}l_{2}}(r')P_{n_{4}\ell_{4}}(r')dr'dr.
\end{eqnarray}

Here, $r_{>}$ ($r_{<}$) is the larger (smaller) of the two radial parameters $r$ and
$r'$.

The $F^{\lambda }$ weight factors for the direct Coulomb interaction   for open atomic shells are
tabulated for all terms in \cite{nk}. The $F^{\mu }$ weight factors for the direct Coulomb interaction as well as the  $G^{\nu }$  weight factors for the `exchange Coulomb interaction that are presented  in  (\ref{uc}) depend on the values of $LS$
and $L'S'$ for the  interacting shells $k$ and $p$ as well as on
the total values of the $L_{t}$  angular and $S_{t}$ spin moments \cite{sob}.

For the closed shell atom with  $N=2(2\ell+1)$ electrons in the shell, the weight factors before the Coulomb integrals are
determined as \cite{sob}:

\begin{equation}
F^{0}=\frac{\left( 4\ell+2\right) (4\ell+1)}{2}=C_{\left( 4\ell+2\right) }^{2},
\label{f0}
\end{equation}

whereas

\begin{equation}
F^{\lambda}=-\left( 2\ell+1\right) ^{2}\left(
\begin{array}{ccc}
\ell & \lambda & \ell \\
0 & 0 & 0%
\end{array}%
\right) ^{2}  \label{fk}
\end{equation}

To avoid accounting for the exact angular-momentum-coupling scheme in our study, for the sake of simplicity, in the present work we use the
average-term approximation, the essence of which is based on the introduction of the average direct Coulomb and exchange weight factors
into the interelectron interaction potential, as follows.

Since the expression for $F^{0}$ is proportional to the number of interacting
pairs of electrons in the closed shell of $4\ell+2$ electrons, it is natural to expect
that the same is true for the $F^{\lambda}$ coefficients as well. Thus, for the $N$-electron open shells, we write
Equations~(\ref{f0}) and (\ref{fk}) as follows:

\begin{equation}
F^{0}(\ell^{N})=\frac{N(N-1)}{2}=C_{N}^{2}  \label{f0a}
\end{equation}

\begin{eqnarray}
F_{\rm av}^{\lambda}(l^{N})=-\left( 2\ell+1\right) ^{2}\nonumber \\
\left(
\times\begin{array}{ccc}
\ell & \lambda & \ell \\
0 & 0 & 0%
\end{array}%
\right)^{2}
 \frac{N(N-1)}{\left( 4\ell+2\right) (4\ell+1)}  \label{fka}
\end{eqnarray}

Equation~(\ref{f0a}) for $F^{0}$ is exact for unfilled shells.

The exact $F^{\lambda}$ weight
factors for $\lambda >0$ depend on the values of $L$, $S$, and additional
quantum numbers. These factors are tabulated in \cite{nk} for all possible terms.
The $F_{\rm av}^{\lambda}$
value  equals the value of $F_{LS}^{\lambda}$ averaged over all $LS$
terms of the $\ell^{N}$ configuration using  their $(2S+1)(2L+1)$ statistical weights.

Next, for the interaction of two closed shells, $\ell^{(4\ell+2)}$ and $\ell'^{(4\ell' +2)}$, the equations for the weight factors are as follows:

\begin{equation}
F^{0}(\ell^{4\ell+2},\ell'^{4\ell'+2})=(4\ell+2)((4\ell'+2),
\label{f0t}
\end{equation}

\begin{eqnarray}
G^{\nu }(\ell^{4\ell+2},\ell'^{4\ell'+2})= \nonumber \\
-2\left( 2\ell+1\right)
(2\ell'+1)
 \left(
\begin{array}{ccc}
\ell & \nu  & \ell' \\
0 & 0 & 0%
\end{array}%
\right)^{2}\   \label{gkt}
\end{eqnarray}

For closed shells, $F^{\mu}=0$ for $\mu\neq 0$.

Equation~(\ref{f0t}) indicates that the intershell interaction is proportional to
the numbers of electrons in the interacting shells. This allows us to write the average
weighs factors for a similar interaction between the open shells as follows

\begin{equation}
F^{0}(\ell^{N},\ell'^{N'})=NN',
\end{equation}

and

\begin{eqnarray}
G^{\nu }(l^{4\ell+2},\ell'^{4\ell'+2})=\nonumber \\
\frac{-2\left( 2\ell+1\right)
(2\ell'+1)NN'}{\left( 4\ell+2\right) \left( 4\ell'+2\right) }\left(
\begin{array}{ccc}
\ell & \nu  & \ell' \\
0 & 0 & 0%
\end{array}%
\right) ^{2}\
\end{eqnarray}

 In our study of the structured open-shell fullerene anions, these weight factors were substituted into the Equation~(\ref{uc}) for the interelectron interaction potential to calculate the term-averaged
energies and $P_{n\ell}$ radial parts of the wavefunctions of electrons.

\end{document}